\documentclass[10pt,a4paper]{article}
\usepackage{a4wide}
\usepackage{amsfonts}
\usepackage{amssymb,amsmath}
\usepackage[dvips]{color}
\usepackage{graphicx}
\usepackage{floatflt}

\begin{document}

\title{Relational Description of the Measurement \\
Process in Quantum Field Theory}
\author{Rodolfo Gambini$^{1}$ and Rafael A. Porto$^{ 1}$\\
1. Instituto de F\'{\i}sica, Facultad de Ciencias, \\ Igu\'a 4225,
esq. Mataojo, Montevideo, Uruguay.}
\date{June 8th 2001}

\maketitle

\begin{abstract}

We have recently introduced a realistic, covariant, interpretation
for the reduction process in relativistic quantum mechanics. The
basic problem for a covariant description is the dependence of the
states on the frame within which collapse takes place. A suitable
use of the causal structure of the devices involved in the
measurement process allowed us to introduce a covariant notion for
the collapse of quantum states. However, a fully consistent
description in the relativistic domain requires the extension of
the interpretation to quantum fields. The extension is far from
straightforward. Besides the obvious difficulty of dealing with
the infinite degrees of freedom of the field theory, one has to
analyze the restrictions imposed by causality concerning the
allowed operations in a measurement process. In this paper we
address these issues. We shall show that, in the case
of partial causally connected measurements, our description allows
us to include a wider class of causal operations than the one resulting
from the standard way for computing  conditional probabilities. 
This alternative description could be experimentally tested. A
verification of this proposal would give
a stronger support to the realistic interpretations of the states
in quantum mechanics.
\end{abstract}

\section{Introduction}

In a previous paper, we have introduced a realistic, covariant,
interpretation for the reduction process in relativistic quantum
mechanics. The basic problem for a covariant description is the
dependence of the states on the frame within which collapse takes
place. More specifically, we have extended the tendency
interpretation of standard quantum mechanics to the relativistic
domain. Within this interpretation of standard quantum mechanics,
a quantum state is a real entity that characterizes the
disposition of the system, at a given value of the time, to
produce certain events with certain probabilities. Due to the
uniqueness of the non-relativistic time, once the measurement
devices are specified, the set of alternatives among which the
system chooses is determined without ambiguities. In fact, they
are associated to the properties corresponding to a certain
decomposition of the identity. The evolution of the state is also
perfectly well defined. For instance, if we adopt the Heisenberg
picture, the evolution is given by a sequence of states of
disposition. The dispositions of the system change during the
measurement processes according to the reduction postulate, and
remain unchanged until the next measurement. Of course, the
complete description is covariant under Galilean transformations.

In Ref\cite{Nos1} we proved that a relativistic quantum state may
be considered as a multi-local relational object that
characterizes the disposition of the system for producing certain
events with certain probabilities among a given intrinsic set of
alternatives. A covariant, intrinsic order was introduced by
making use of the partial order of events induced by the causal
structure of the theory. To do that, we have considered an
experimental arrangement of measurement devices, each of them
associated with the measurement of certain property over a
space-like region at a given proper time. No special assumption
was made about the state of motion of each device. Indeed,
different proper times could emerge from this description due to
the different local reference systems of each device. Thus, we may
label each detector in an arbitrary system of coordinates by an
open three-dimensional region $R_a$, and its four-velocity $u_a$.
We now introduce a partial order in the following way: The
instrument $A_{{R_1},{u_1}}$ precedes $A_{{R_2},{u_2}}$ if the
region $R_2$ is contained in the forward light cone of $R_1$. Let
us suppose that ${A^0}_{{R},{u}}$ precedes all the others. Then,
it is possible to introduce a strict order without any reference
to a Lorentz time as follows. Define $S^1$ as the set of
instruments that are preceded only by $A^0$.  Define $S^2$ as the
set of instruments that are preceded only by the set $S^1$ and
$A^0$. In general, define $S^i$ as the set of instruments that are
preceded by the sets $S^j$ with $j <i$ and $A^0$. The crucial
observation is that all the measurements on $S^i$ can be
considered as "simultaneous". In fact, they are associated with
local measurements performed by each device, and hence represented
by a set of commuting operators. As the projectors commute and are
self-adjoint on a ``simultaneous" set $S^i$, all of them can be
diagonalized on a single option. These conditions ensure that the
quantum system has a well defined disposition with respect to the
different alternatives of the set $S^i$. In other words one can
unambiguously assign conditional probabilities after each
measurement for the events associated to the set $S^i$.

In relativistic quantum mechanics, this description is
only
consistent up to lambda Compton corrections. In
fact, the
corresponding local projectors exist and
commute, up to Compton
wavelengths \cite{Nos1,Nos2,MaRo}.
A fully consistent description of the
measurement process
in the relativistic domain requires the
extension of the
interpretation to quantum fields.This extension
is far from
trivial. Besides the obvious difficulty of dealing
with the
infinite degrees of freedom of the field theory, one has
to
face some issues related with the lack of a covariant notion of
time order of the quantum measurements. In fact, there is not a
well defined description for the Schroedinger evolution of the
states on arbitrary foliations of space time, even for the free
scalar quantum field in a Minkowski background.\cite{Torre}
Although
the evolution is well defined in the Heisenberg picture,
in general
the operators associated with global space-time
foliations are not
self-adjoint. It is not guaranteed that in the particular case of the
field operators this problem will appear. However, it is clear that a
careful treatment is required in order to insure that they are well
defined operators.

Another issue concerns the causal restrictions on the observable
character of certain operators in Q.F.T. As it has been shown by
many authors, causality imposes further restrictions on the
allowed ideal operations on a measurement process. This
observation arise when one considers some particular arrangements
composed by partial causally connected measurements. It has been
shown that while some operators are admissible in the relativistic
domain, many others are not allowed by the standard
formalism\cite{A-A,Sorkin,Preskill,Preskill2}. Although this
conclusion is correct, it is based on standard Bloch's notion for
ordering the events in the relativistic domain. Remember that
Bloch's approach consists on taking any Lorentzian reference
system and hence:{\it "...the right way to predict results
obtained at $C$ is to use the time order that the three regions
$A,B,C$ have in the Lorentz frame that one happens to be
using"}\cite{Bloch}. Nevertheless, we have introduced in
\cite{Nos1} another covariant notion of partial order. Though both orders
coincides in many cases, they imply different predictions for the
cases of partial causally connected measurements. Here we shall
show that our notion of intrinsic order allows us to extend the
allowed causal operators to a wider and natural class.

In this paper we will consider the explicit case of a free, real,
scalar field in a Minkowski space-time. The field operators
smeared with local smooth functions are quantum observables
associated with ideal measurement devices. They are associated to
projectors corresponding to different values of the observed
fields. We shall prove that the projectors associated with
different regions of the $S^i$ option commute. This allows us to
extend the real tendency interpretation to the quantum field
theory domain giving a covariant description of the evolution of
the states in the Heisenberg picture. As in relativistic quantum
mechanics, the states are multi-local relational objects that
characterize the disposition of the system for producing certain
events with certain probabilities among a particular an intrinsic
set of alternatives. The resulting picture of the multi-local and
relational nature of quantum reality is even more intriguing than
in the case of the relativistic particle. We shall show that it
implies a modification of the standard expression for conditional
probabilities in the case of partial causally connected
measurements, allowing to include a wider range of causal
operators. Our description could be experimentally tested. A
verification of our predictions would give a stronger support to
the realistic interpretations of the states in quantum mechanics.

The paper is organized as follows: In section 2, we develop our
approach for a real free scalar field showing that it is possible
to give a standard description of the measurement process of a
quantum field. In section 3, we show that this approach is
consistent with causality and provides predictions for conditional
probabilities that differ from the standard predictions in the
case of partial causally connected measurements. We also discuss
the resulting relational interpretation of the quantum world. We
present some concluding remarks in Section 4. The existence of the
projectors as distributional operators acting on the Fock space is
discussed in the Appendix.

\section{The free K-G field}

We shall study the relational tendency theory of a real free K-G
field, evolving on a flat space-time. We start by considering the
experimental arrangement of measurement devices,
${A^f}_{{R_a},{u_a}}$ each of them associated with the measurement
of the average field

\begin{equation} {{\Phi}^a} \equiv
{\Phi^{f^a}}_{R_a}(t^{L_a})=\int_{R_a}{f^a(y)\Phi(t^{L_a},y)d^3y}
\end{equation}

where $f^a$ is a smooth smearing function with compact support
such that it is non-zero in the $R_a$ region associated with the
instrument that measures the field. The decomposition
$(t^{L_a},x^j)$ corresponds to the coordinates in the local
Lorentz rest frame of the measurement device located in $R_a$.

The scalar field operators satisfy the field equations

\begin{equation}
({\eta}^{{\mu}{\nu}}{\nabla}_{\mu}{\nabla}_{\nu}-m^2){\Phi}=0
\end{equation}

and the canonical commutation relations:

\begin{equation}
[{\Phi}(t^{L_a},y^j),{\Pi}(t^{L_a},x^j)]=i{\delta}(x^j-y^j)
\end{equation}

\begin{equation} [\Phi (t^{L_a},x^j),\Phi(t^{L_a},y^j)]= 0 \end{equation}

Thus, we may write the field operator in terms of its Fourier
components as follows:

\begin{equation} \Phi(x^{\mu})={\int} d^3k[{\bf {a}}(k) g_k(x^{\mu})+{\bf
{a}}^{\dagger}(k){g_k}^{\star}(x^{\mu})] \end{equation}

with

\begin{equation} g_k=\frac{1}{\sqrt{{(2{\pi})^3}2{\omega}_k}}
exp(-i{k^{\nu}}x_{\nu}) \end{equation}

and

\begin{equation} k_0={\omega}_k=+{\sqrt{{k^{j}}{k_{j}}+m^2}}
\end{equation}

Generically, the devices belonging to the same set of alternatives
$S^i$ will lie on several spatially separated non-simultaneous
regions. Thus, in order to describe the whole set of alternatives
in a single covariant Hilbert space ${\cal H}$ we will have to
transform these operators to an arbitrary Lorentzian coordinate
system. We shall exclude  accelerated detectors, and consequently,
we will have an unique decomposition of the fields in positive and
negative frequency modes. This procedure allows us to define the
Hilbert space in the Heisenberg picture on any global Lorentz
coordinate system. The crucial observation is that all the
measurements on $S^i$ can be considered as "simultaneous". In
fact, two arbitrary devices of $S^i$ are separated by space-like
intervals, and therefore, we shall prove that the corresponding
operators ${\Phi^a}$, represented on ${\cal H}$, commute. What
remains to be proved is that they are unbounded self-adjoint
operators in the Fock space ${\cal F}$ of the scalar field and
therefore they can be associated with ideal measurements. A
measurement will produce events on the devices belonging to $S^i$
and the state of the field will collapse to the projected state
associated to the set of outcomes of the measurement. The
determination of the corresponding projectors is a crucial step of
our construction. We are also going to prove that the construction
is totally covariant and only depends on the quantum system, that
is, the scalar field and the set of measurement devices. All the
local operators $\Phi^a$ are represented on a generic Hilbert
space via boosts transformations, and the physical predictions are
independent on the particular space-like surface chosen for the
definition of the inner product. Notice that we are not filling
the whole space-time with devices. Instead we are considering a
set of local measurements covering partial regions of space-time.
If we had chosen the first point of view, we would run into
troubles. Indeed, it was shown \cite{Torre} that the functional
evolution cannot be globally and unitarily implemented except for
isometric foliations.

Once the projectors in the local reference frame of each detector
has been defined, we need to transform them to a common, generic,
Lorentz frame where all the projectors will be simultaneously
defined. In other words, recalling that Hilbert spaces
corresponding to two inertial systems of coordinates are unitarily
equivalent we will represent all the projectors on the same space.
The projectors and the smeared field operators transform in the
same way, that is:

\begin{equation} {\Phi^a}_L= U^{-1}(L,L_a){{\Phi}^a}U(L,L_a)
\end{equation}

where $U(L,L_a)$ is the unitary operator related to the boost
connecting the generic Lorentz frame with the local frame of each
device. Since we are dealing with the Heisenberg picture, the
states do not evolve, and only the operators change with time. One
can parameterize the evolution, with the time in the  local reference 
frame of the device located in the region $R_a$, 
or what is equivalent with the proper time associated to this device. 
The projectors corresponding to the observation of a given value of 
the field
${\Phi}^a$ at a given proper time may be represented on the
Hilbert space associated with any Lorentz frame.

We are now ready to study the spectral decomposition of the
$\Phi^a$ operators. We start by solving the eigenvalue problem in
the field representation. We shall work in the proper reference
system where the measurement device is at rest. We shall proceed
as follows, we start by choosing the field polarization and
defining the Fock space. Then we shall determine the eigenvectors
of the quantum observables $\Phi^a$, and show that they are well
defined elements of  the Fock space.

On this representation the field operators are diagonal and the
canonical momenta are derivative operators,

\begin{equation}
 <\phi|\Phi(x^j)|\Psi>=\phi(x^j)\Psi[{\phi}], \end{equation}

\begin{equation}
 <\phi|\Pi(x^j)|\Psi>= \frac{1}{i} \frac{\delta}{\delta\phi(x^j)}
\Psi[{\phi}].
 \end{equation}

The inner product is given by: \footnote{We are working in the
functional
Schroedinger representation\cite{Jackiw} which is convenient because of 
its close analogy with quantum mechanics. A more rigorous presentation 
would require the introduction of a Gaussian  or a White Noise measure 
in the infinite dimensional space.\cite{MoThiVe}} 

\begin{equation}
 <\Psi(\phi)|\Gamma(\phi)>=\int D\phi{\Psi}^*(\phi)\Gamma(\phi)
\end{equation}

and the eigenvectors of the field operators $|\phi>$ satisfy

\begin{equation} <{\phi}|\tilde{\phi}>={\delta}(\phi-\tilde{\phi})
\end{equation}

The fields transform as scalars under Lorentz transformations and
the inner product is Lorentz invariant.

Let us now proceed to the construction of the Hamiltonian and the
vacuum state in this representation.

The Hamiltonian operator is

\begin{equation}
 H= \int_{\Sigma_0}{\frac{1}{2}[{\Pi^2} +\Phi{\hat
\omega}^2\Phi}]\label{ham} \end{equation}

where

\begin{equation} {\hat \omega}^2={-{\nabla}}^2+m^2 \end{equation}

The functional equation for the vacuum state, $\Psi_0[{\phi}]$,
turns out to be:

\begin{equation} \frac{1}{2}
 \left(\int_{\Sigma_0}{-\frac{{\delta}^2}{{\delta}\phi{\delta}\phi}}
+\phi{\omega}^2{\phi}\right) {\Psi_0}(\phi)= E_0 {\Psi_0}(\phi) .
\end{equation}

The vacuum solution is of the form

\begin{equation} \Psi_0[\phi]= det^{1/4}(\frac{\omega}{\pi}) exp\left({-1
\over
2}\int_{\Sigma_0}d^3zd^3x\phi(x^j){\omega}(z^j-x^j)\phi(z^j)\right)
\end{equation}

where

\begin{equation}
{\omega}(x^j-y^j)=\sqrt{(-{{\nabla^2}_{x^j}}+m^2)}\delta(x^j-y^j)
\equiv \int d^jk w_k e^{ik^j(x^j - y^j)}
 \end{equation}

It can be easily seen that the energy of the ground state turns
out to be $E_0=\frac{1}{2}tr{\omega}$ which diverges due to the
zero mode contributions. The normalization factor also becomes
infinite due to ultraviolet divergences. As it is well known and
we shall show in what follows these infinities are harmless.

It is easy to show that the ground state is annihilated by:

\begin{equation} a(x^j)= \frac{1}{\sqrt{2}}\int_{\Sigma_0} d^3y
{\omega}^{1/2}(x^j-y^j)\Phi(y^j)
+i{\omega}^{-1/2}(x^j-y^j)\Pi(y^j).
\end{equation}

The Hamiltonian operator (\ref{ham}) is not well ordered, but its
well ordered form may be immediately obtained by subtracting
$E_0$. The corresponding creation operator may be immediately
defined. The action of the Fourier component of the creation
operator for the $k$ mode on the vaccuum state leads to the state
$\Psi_1$ with eigenvalue $E_1=w_k+E_0$. The set of functional
states given by the repeated action of the creation operator
defines a basis of the Fock space. This is the orthonormal basis
of "functional Hermite polynomials" \cite{Jackiw}.

Now, we proceed to define the field operator in the proper Lorentz
system. The Heisenberg equation for the field will read:

\begin{equation} i{{\partial}\Phi \over {\partial}t} = [\Phi,H]
\end{equation}

\begin{equation} i{{\partial}\Pi \over {\partial}t} = [\Pi,H]
\end{equation}

That is:

\begin{equation} i{{\partial}\Phi \over {\partial}t} = \Pi \end{equation}

\begin{equation} i{{\partial}\Pi \over {\partial}t} = {\omega}^2\Phi
\end{equation}

that leads to the K-G equation for the field operator. The general
solution of this equation turns out to be:

\begin{eqnarray} \phi(x) &=& \int_{\Sigma_0}\phi(y)
\stackrel{\leftrightarrow}{\partial_n} D(x-y)dV\nonumber\\
&=&\int_{\Sigma^0}\phi(y) \stackrel{\leftrightarrow}{\partial_\mu}
D(x-y)d\sigma_\mu \label{uno} \end{eqnarray}

where $D$ is the homogeneous antisymmetric Klein-Gordon
propagator. It can be easily seen, by making use of the Gauss
Theorem, that this integral does not depend on the space-like
surface. Thus, we can choose $\Sigma_0$ as the initial surface
$y^0=0$ of the proper Lorentz system. It is easy to define, from
this expression, the evolving operator. In fact, the solution
depends on the value of $\phi$ and its temporal derivate, which is
just its conjugate momentum $\Pi$, over $\Sigma_0$. Then the field
will read:

\begin{equation} \Phi(x^j,t^{L_a})= \int_{\Sigma_0} dy^3
\phi(y^j){\partial}_{y^0} D(x^{\mu},y^j)-D(x^{\mu},y^j){1 \over
i}\frac{\delta} {\delta\phi(y^j)} \end{equation}

We still have to show  that the projectors corresponding to a
measurement of the smeared field exist. More precisely, we will
show that its action is well defined in the Hilbert space. We
start with the determination of the eigenstates corresponding to
the eigenvalue $\phi_E$ of the field operator $\Phi(x^j,t^{L_a})$
on the proper system.

\begin{equation} <\phi|\phi_E,t^{L_a}>= C e^{{i \over 2}<{\phi}A{\phi}>}
e^{{i \over 2}<\phi_E A \phi_E>} e^{i<B{\phi_E}{\phi}>}
\end{equation}

where $A$ and $B$ are propagators with Fourier components given
by:

\begin{eqnarray} A(p,t^{L_a}) \equiv
\frac{\omega_{p}}{tan({\omega}_{p}t^{L_a})}\\ B(p,t^{L_a}) \equiv
\frac{\omega_{p}}{sin({\omega}_{p}t^{L_a})} \end{eqnarray}

and we have used brackets for representing that the fields are
integrated on $\Sigma_0$.

The normalization factor is determined by imposing the
orthonormality of the eigenstates,

\begin{equation} <\phi_E, t^{L_a}|\phi_E',t^{L_a}>= |C|^2 e^{{-i \over
2}<{\phi}_E A{\phi}_E >}e^{{i \over 2}<{\phi}_E' A
{\phi}_E'>}{\prod_{p} 2\pi{sin({\omega}_{p}t^{L_a})
\over{\omega}_p}} \delta(\phi_{E }(p)-\phi_{E'}(p)) \end{equation}

and it is given by:

\begin{equation} C=\left(\prod_{p}{\frac{\omega_p}{{2\pi}i
sin(\omega_{p}t^{L_a})}}\right)^{1 \over 2} \end{equation}

We also notice the interesting fact that if we take the limit:

\begin{equation} <\phi|\phi_{E},t^{L_a}\rightarrow 0>=
\left(\sqrt{\frac{1}{{2\pi}i t^{L_a}}}\right)^n \exp {i \over
2}\left(\frac{(<\phi-\phi_E>)^2}{t^{L_a}}\right) \end{equation}

One recognizes the free propagator, and the delta function
$\delta(\phi-\phi_E)$ as one could expect.

One needs to introduce an infrared and ultraviolet regularization.
The infrared regularization may be implemented by defining the
fields in a periodic box. This allows to have a well defined
normalization factor. The box will break the Lorentz invariance,
but as we are dealing with local measurements, and we take the
sides of the box much larger than the local region under study,
this fact does not have any observable effect. We shall discuss
the ultraviolet regularization later.

Let us now smear the field operators with smooth functions in
order to have well defined eigenvectors. The smeared fields are
the relational quantities that will be actually measured. Let us
call $\phi_\Delta$ one of the eigenvalues that gives the real
quantity $\Delta$ when smeared with the function $f^a$. That is,
$\int_{R_a}f^a(y^j)\phi_\Delta(y^j)dy^j=\Delta$. Thus, $\Delta$ is
the outcome of the relational observation. Let us denote the
corresponding eigenstate $|\phi_\Delta,t^{L_a}>$

Now we are ready for defining the projector for the $R_a$ region.
It is given by: \begin{equation}
<\phi|{P^a}_{F_\Delta}(t^{L_a})|\phi'>=\int_{F_\Delta} d\Delta
\int{d\phi_{\Delta}<\phi|\phi_{\Delta},t^{L_a}>
<\phi_{\Delta},t^{L_a}|\phi'>\delta(<f^a\phi_\Delta>-\Delta)}
\end{equation}

Where $F_\Delta$ is a partition of the possible values of
$\Delta$.\footnote{For definiteness it it necessary to divide the
real line in disjoint intervals. Therefore the regions $F_\Delta$
are open subsets of ${\cal R}$} Notice that all the integrals are
over the surface $\Sigma_0$ since $f^a$ has compact support in
$R_a$. \footnote{Recall that the field are given in a Heisenberg
picture where the operators evolve respect to those living in the
initial data. Therefore we can take the region $R_a$ as part of
$\Sigma_0$ since we are in the proper Lorentz system where the
device is at rest}

Furthermore:

\begin{eqnarray} &<\phi|{P^a}_{F_\Delta}(t^{L_a})|\phi'>=&\nonumber\\
&\int_{F_\Delta}d\Delta e^{{-i \over 2}<{\phi}A{\phi}>}e^{{i
\over2}<\phi' A \phi'>} \int d\alpha e^{-i\alpha \Delta} \prod_{p}
\delta \left(\phi'(p)-\phi(p)+\alpha \frac{sin(\omega_{p}
t^{L_a})}{\omega_{p}}f^a(p)\right)&
\end{eqnarray}

which is a functional distribution over the Hilbert space
$L^2[d\phi]$, once the infrared regularization is taken into
account.

For instance, if we compute its matrix elements among two vectors
of the Hilbert space:

\begin{eqnarray}
&<\Psi(\phi)|{P^a}_{F_\Delta}(t^{L_a})|\Gamma(\phi)>=&\nonumber\\
&\int_{F_\Delta} d\Delta \int d\alpha e^{-i\alpha\Delta}\int
D\phi_p \Psi\left(\phi(p)+\alpha\frac{sin(\omega_{p}
t^{L_a})}{\omega_{p}}f^a(p)\right)\Gamma(\phi(p))\times &
\nonumber\\ & e^{2i\alpha{\left(\frac{2\pi}{L}\right)^3}{\sum}_p
f^a(p)cos(\omega_p t^{L_a})\phi(p)}
e^{i\alpha^2{\left(\frac{2\pi}{L}\right)}^3 {\sum}_p
f^a(p)\frac{sin(\omega_p t^{L_a})cos(\omega_p
t^{L_a})}{{\omega_p}}f^a(p)}&
\end{eqnarray}

where $L^3$ is the volume of the box where we may put the field in
order of avoiding the infrared divergences.\\

In order to prove that ${P^a}_{F_\Delta}(t^{L_a})$ is a projector
we start by observing that:

\begin{equation} \int {d\phi_{E}<\phi|\phi_{E},t^{L_a}>
<\phi_E,t^{L_a}|\phi'>}= \delta(\phi-\phi') \end{equation}

This property allows us to construct a decomposition of the
identity for a set of projectors associated to open portions
$F_\Delta$ of the reals ${\cal R}$ such that $F_\Delta \cap
F_{\Delta'}=0$, and $\bigcup F_\Delta \sim {\cal R}$ up to a zero
measure set. Therefore \footnote{This is reminiscent of the
decomposition of the position operator in standard quantum
mechanics in terms of open intervals that satisfy $\sum_{\Delta x}
\int_{\Delta x} dx <f|x><x|f'>= <f|f'>$}:

\begin{equation} \sum_{ F_\Delta}
<\Psi(\phi)|{P^a}_{F_\Delta}|\Gamma(\phi)>=
<\Psi(\phi)|\Gamma(\phi)>
\end{equation}

Furthermore, if ${\tilde \Delta} \in F_\Delta$:

\begin{equation} {P^a}_{F_\Delta}(t^{L_a})|\phi_{\tilde
\Delta},t^{L_a}>=|\phi_{\tilde \Delta},t^{L_a}> \end{equation}

If not: \begin{equation} {P^a}_{F_\Delta}(t^{L_a})|\phi_{\tilde
\Delta},t^{L_a}>=0 \end{equation}

Finally two projectors associated to different spatial regions
commute. This is a consequence of the commutation of the local
operators $\Phi^a$. Indeed, if the local regions, $R_a$, $R_b$
define the same proper Lorentz frame the commutation of $\Phi^a$
and $\Phi^b$ for space-like separation is straightforward and
hence the projectors commute. If the regions are not simultaneous,
one needs to transform both operators to a common Lorentz frame.
Let us call $U_{ab}$, the Lorentz transformation connecting both
regions, then the relevant commutator will be
$[{U_{ab}}^{\dagger}\Phi^bU_{ab},\Phi^a]$. As an arbitrary Lorentz
boost may be written as a product of infinitesimal
transformations, $U_{ab}=Id+{\epsilon}^{{\mu}{\nu}}M_{{\mu}{\nu}}$
it is sufficient to consider:

\begin{equation} {U_{ab}}^{-1}\Phi^bU_{ab}=
\Phi^b+{\epsilon}^{{\mu}{\nu}}[M_{{\mu}{\nu}},\Phi^b]
\end{equation}

The commutator of $M_{{\mu}{\nu}}$ with $\Phi^a$ only involve
canonical operators evaluated at points of the region $R_a$, their
commutator with operators associated to a region $R_b$ separated
by a space-like interval will commute.

Thus, also for non-simultaneous regions, space-like separated
projectors commute. In the Appendix we prove that this projectors
have a well defined action on the Fock space of the free Klein
Gordon field.

Thus the projectors associated to different local measurement
devices on a ``simultaneous" set $S^i$ commute and are
self-adjoint. These properties insure that they can be
diagonalized on a single set of alternatives, and the quantum
system has a well defined, dispositional, state with respect to
the different alternatives of the set $S^i$. In the Heisenberg
picture, the evolution is given by a sequence of states of
disposition. The dispositions of the system change during the
measurement processes according to the reduction postulate, and
remain unchanged until the next measurement. As in the case of
relativistic quantum mechanics, the system provides, in each
measurement, a result in devices that may be located on arbitrary
space-like surfaces. Notice that contrary to what happens with the
standard Lorentz dependent description of the reduction process,
here the conditional probabilities of further measurements are
unique. It is in that sense that the dispositions of the state to
produce further results have an objective character.

\section{Causality vs. the Intrinsic Order}

As we mentioned before, it has been recently observed by many
authors \cite{A-A,Sorkin,Preskill,Preskill2} that the standard
time order of ideal measurements in a Hilbert space may imply
causal violations if partially connected regions are taken into
account. Here we shall show that although this analysis is
correct, it is based on a different notion for the ordering of the
events. If one defines the partial order as we did in Ref
\cite{Nos1}, one may extend the causal predictions of the theory,
and the reduction process is covariant and consistent with
causality for a wide and natural class of operators.\\

Let us suppose, following Sorkin \cite{Sorkin}, that the devices
performing the observation are not completely contained in the
light cones coming from the previous set. We are therefore,
interested in the case where only a portion of certain instrument
is contained inside the light cone of the previous set. We could
generalize the previously introduced notion of order by saying
that $B$ follows the instrument $A$ if at least a portion of $B$
lies inside the forward light cone coming from $A$. With this
ordering, let us to consider a particular arrangement for a set of
instruments which measure a particular observable on a
relativistic quantum system. Suppose three local regions: $A$,
$B$, $C$ with their corresponding Heisenberg projectors:
${P^A}_a$, ${P^B}_b$, ${P^B}_c$ associated to values of certain
Heisenberg observables over each region. We arrange the regions
such that some points of $B$ follows $A$ and some points of $C$
follows $B$ but $A$ and $C$ are spatially separated(see figure
1)\cite{Sorkin}. It is easy to build such arrangement, even with
local regions. In this context, due to microcausality, the
commutation relations between the observables and the projectors
will be:

\begin{equation}
[{P^A}_a,{P^B}_b]\neq 0
\end{equation}
\begin{equation}
[{P^B}_b,{P^C}_c]\neq 0
\end{equation}
\begin{equation}
[{P^A}_a,{P^C}_c]=0
\end{equation}

Let us suppose that one uses this new notion of order, to define
the sequence of options $S^1$,$S^2$, $S^3$ and the corresponding
reduction processes followed by a quantum system. Then, since the
new order implies $A<B<C$, one immediately notices that the $A$
measurement affects the $B$ measurement and also the $B$
measurement affects the $C$ measurements. Consequently, one should
expect that the $A$ measurement would affect the $C$ measurement,
leading to information traveling faster than light between $A$ and
$C$, which are space-like separated regions.

\begin{figure}[htbp]
\begin{center}
\includegraphics[width=6cm]{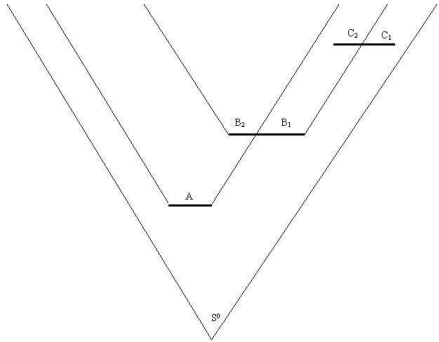}
\caption{Sorkin's arrangement with partial causally connected
local regions.}
\end{center}
\end{figure}

One could immediately prove this fact as follows, let us suppose
that the state of the field was prepared by a initial measurement,
that precedes the whole arrangement, whose density operator we
denote by $\rho_0$. Now, the probability of having the result
$a,b,c$ in the regions $A,B,C$ given the initial state $\rho_0$
is, using Wigner's formula:

\begin{equation}
{\cal P}(a,b,c|\rho_0)=Tr[{P^C}_c{P^B}_b{P^A}_a{\rho}_0{P^A}_a{P^B}_b]
\label{probm}
\end{equation}

This is the standard result that we would have obtained by making use of
Bloch's notion of order.
Thus, one notices that an observer located in $C$ could know with
certainty if a measurement has been performed by $A$. In fact,
assume that a non-selective measurement has occurred on the region
$A$, and ask for the probability of having $c$ under this
hypothesis.

Then, one arrive to the probability:

\begin{equation}
{\cal P}(\mathrm{unknown} \, a,b,c|\rho_0)=
{{\sum}_a}\mathrm{Tr}[{P^C}_c{P^B}_b{P^A}_a{\rho}_0
{P^A}_a{P^B}_b]
\end{equation}

One immediately notices that this probability depends on whether
the $A$ measurement was carried out or not, independently of the
result. This is due the non-commutativity of the projector $P_A$
with $P_B$, and $P_B$ with $P_C$, that prevents us for using the
identity ${\sum}_a{P^A}_a=Id$.\\

Notice that we have assumed that the $B$ measurement is known.
However, since the $B$ region is partially connected with region
$C$, a portion of $B$ will not be causally connected with $C$ and
therefore, the preservation of causality would require that the
measurement carried out by $B$ should be taken as non-selective
respect to an observer localized on $C$.\footnote{The resulting
value on the measurement carried out on $B$ can not be transmitted
causally to an observer in $C$.}

If we take this fact into account one can prove that even with a
non-selective measurement on $B$ one arrive to causal problems. In
fact, we notice that the probability of having $c$, no matter

the result on $A,B$ is:

\begin{equation}
{\cal P}(\mathrm{unknown}\; a,\mathrm{unknown}\; b,c|\rho_0)=
\sum_a\sum_b {\mathrm{Tr}} [{P^C}_c{P^B}_b{P^A}_a{\rho}_0
{P^A}_a{P^B}_b]\label{probm2},
\end{equation}

which depends on whether or not the $A$ and $B$ measurements were
carried out. Hence, if one starts from a different definition of
the partial ordering of the alternatives, in terms of a partial
causal connection, one gets faster than light signals for a wide
class of operators which prevent us to eliminate the $A$
measurement. In those cases the observer $C$ could know with
certainty if the previous two measurements were carried out or
not. There is not any violation with respect to the $B$
observation since an observer at $C$ may be causally informed
about a measurement carried out at $B$. However, the above
analysis implies faster than light communication with respect to
the $A$ measurement since it is space-like separated from $C$.
Therefore, the requirement of causality strongly restricts the
allowed observable quantities in relativistic quantum mechanics.\\

In what follows we are going to show that our description is
consistent with causality for a wider range of operations. The key
observation is that our notion of partial ordering requires to
consider the instruments as composed of several parts each one
associated to different measurement processes. That is, in the
case where only a portion of the instrument is causally connected,
one needs to decompose the devices in parts such that each part is
completely inside (or outside) the forward light cone coming from
the previous devices. Now the alternatives belonging to one option
$S^i$ are composed by several parts of different instruments. In
fact, a particular device could contain parts belonging to
different options. Although the measurement performed by any
device is seen as Lorentz simultaneous for any local reference
system it will be associated to several events. \footnote{In the
context of Q.F.T. we define an event as the projection of the
state. This generalization is natural since in Q.F.T. one can
associate a negative result with a zero value of certain physical
observable as, for instance, the charge of the field.}

Let us reconsider the previous example with our notion of order
(see figure 1). Let us start with $S^0$ and the preparation of the
state in $\rho_0$. We will call ($B_1$) the part of $B$
non-casually connected with $A$ and ($C_1$) the part of $C$ non
causally connected with $B$. The part of $B$ causally connected
with $A$ ($B_2$) and the part of $C$ causally connected with $B$
($C_2$). Now we can construct the set of options as,
$S^1=(A,B_1,C_1)$ then $S^2=(B_2,C_2)$.

Thus, we need to deal with partial observations. Let us consider
the case where the operator associated to the measurement carried out
on $C$ may be taken as
composed by two partial operators associated to $C_1$ and $C_2$, We
shall denote
the respective eigenvalues as $c_1$ and $c_2$. Notice that, the
individuality of the device still persists since we do not have
access to each result but only to the total result $c$ obtained on
$C$ after the observation. Now it is important to consider how one
gets $c$ through $c_1$ and $c_2$. Let us assume that the result
$c$ is extensive in the sense that $c=f(c_1,c_2)$. This relation
depends on the particular observation we are performing on each
alternative. For instance, let us call $(O^1,O^2)$ the local
operators associated to the observations on $(C_1,C_2)$ and $O$
the operator associated to $C$. Therefore, $f(O^1,O^2)=O$ is the
functional relation between them. For the case of the field
measurements, we will have $f\equiv c_1+c_2$ which is just the
relation $\Phi^{C_1}+\Phi^{C_2}=\Phi^C$. Notice however, that this
hypothesis also includes a wide range of observables. Indeed, it
allows us to measure local operators which involve products of
multiple smeared fields. These operators will imply indeed a non
linear behavior for the functional relation $f(c_1,c_2)=c$.
Now we can compute the probability of observing $c$ for selective
measurements in $A,B$ given the initial state $\rho_0$. In first
place, we have to deal with the measurement of $b$ occurring on
$B$. As we have divided the device in two portions, this result
will be composed by two unknown measurements $b_1$ and $b_2$, such
that $b=f(b_1,b_2)$. Analogously for the probability of having $c$
since it results from two independent measurements in $C_1$ and
$C_2$. Thus, we will have:

\begin{eqnarray}
&{\cal
P}(a,b,c|\rho_0)={\sum}_{(c_1,c_2,b_1,b_2)}\delta(c-f(c_1,c_2))
\delta(b-f(b_1,b_2))\times\\ &\times
Tr[{P^{C_2}}_{c_2}{P^{B_2}}_{b_2}{P^{B_1}}_{b_1}{P^{C_1}}_{c_1}
{P^A}_a{\rho}_0{P^A}_a{P^{C_1}}_{c_1}{P^{B_1}}_{b_1}{P^{B_2}}_{b_2}])=
\nonumber\\&\sum_{b_1}\sum_{b_2}\delta(b-f(b_1,b_2))
Tr[{P^C}_c{P^{B_2}}_{b_2}{P^{B_1}}_{b_1}
{P^A}_a{\rho}_0{P^A}_a{P^{B_1}}_{b_1}{P^{B_2}}_{b_2}]\nonumber
\label{prob1}
\end{eqnarray}

Where we have taken into account that, due to  microcausality:

\begin{equation}
{P^C}_{c}=\sum_{c_1}\sum_{c_2}\delta(c-f(c_1,c_2))
{P^{C_1}}_{c_1}{P^{C_2}}_{c_2} \label{proj}
\end{equation}

The sum on $b_1,b_2$ goes over the complete set of possible
results. The same applies for the $C$ measurement.\\ Now, in order
to study the causal implications we need to compute the
probability of having $c$ for non-selective measurements on $A,B$.
Therefore, one gets:

\begin{eqnarray}
&{\cal P}(\mathrm{unknown}\; a,\mathrm{unknown}\; b,c|\rho_0)=
\sum_a\sum_b{\cal P}(c,a,b|\rho_0)=&
\label{prob2}\\&{\sum}_{(c_1,c_2,b_1)}\delta(c- f(c_1,c_2))
Tr[{P^{C_2}}_{c_2}{P^{B_1}}_{b_1}{P^{C_1}}_{c_1}
{\rho}_0{P^{C_1}}_{c_1}{P^{B_1}}_{b_1}] =
\sum_{b_1}Tr[{P^C}_c{P^{B_1}}_{b_1}\rho_0{P^{B_1}}_{b_1}]&
\nonumber
\end{eqnarray}

Where we have used that $\sum_{b_2}{P^{B_2}}_{b_2}=Id$. Thus, this
probability does not depend on the $A$ measurement and our
description does not lead to any violation of causality during the
measurement process.

Although there is some kind of correlation introduced by the
causally connected part of $B$ with $C$, we will not have any
information about the actual observation made on $B$, as we
noticed before. This correlation is very interesting and could be
experimentally tested. Notice that only the assignment of
probabilities given by equations (\ref{prob1},\ref{prob2}) is
consistent with causality for the general kind of measurements
that we have considered.\\

Several issues concerning the relational interpretation can be
read from the previous analysis. The devices never lose their
individuality as instruments of measurement of a certain
observable, for instance, the local field on certain region.
However what is quite surprising is that while the devices are
turned on for a local proper time $T$, the {\it "decision"} made
by the quantum system with respect to this region is taken by two
non-simultaneous processes within the given intrinsic order. The
local time of measurement is quite different to the internal order
for which the {\it "decisions"} were taken. Now, the set of
"simultaneous" alternatives is composed by portions of several
devices. The individuality of each device is preserved, since we
do not have access to the results of these partial alternatives.
What we observe in each experiment is the total result registered
by each device.\\ Another consequence of our approach concerns the
causal connection among alternatives belonging to different sets
$S^j$. As we have shown, there is correlation among the causally
connected portions of different devices, nevertheless this
correlation does not imply any incompatibility with causality.\\
All these features show a global aspect of the relational tendency
interpretation which is very interesting since the decomposition
is produced by the global configuration of the measurement devices
evolving in a Minkowski space-time, without any reference to a
particular Lorentz foliation.\\

We have considered a measurement arrangement which is reminiscent
to the observation of a non-local property \cite{A-A}.\footnote{Notice
that the operators
associated to each local region may be non-local operators. For
instance the non-linear operator
$\int_{R_a}\int_{R_a}f^a(x)f^a(y)\phi(x)\phi(y)dxdy$ is non local with
respect
to $R_a$.} 
We can indeed naturally extend our
approach to the case of widely separated non-local measurements,
or even widely extended observation.
Now the partial causal connection is simply implemented
taken Sorkin's arrangement, on figure 1, modified to the case of
measurements carried out on disconnected regions $B,C$, or even a
space-like surface.\footnote{One should be careful on extending to
a space-like surface. As we have mentioned it is not possible in
general to introduce a well defined self-adjoint operator
associated to an arbitrary space-like hyper-surface. However, as
we defined the set of alternatives $S^j$, the space-like surface
we may consider will be a portion of a constant time surface on the
Lorentz rest frame of the devices involved in the non-local
measurement. In these cases, it is possible to show that the
relational observable is a well defined self-adjoint operator.}
\footnote{Notice that Sorkin's arrangement is quite natural for
studying the causal implication of the theory. In fact, the
example given by Sorkin in Ref \cite{Sorkin} is indeed a non local
measurement carried out on a spacelike surface. In those cases, of
a widely extended non-local measurement, the partial connection is
always fulfilled.} The conclusion is the same. In the cases of
partial causally connected measurements our description includes a
wider range of causal operators than the standard
approach.\footnote{It is important to remark that in our case, due
to Microcausality, the standard expression (\ref{probm}) is causal
for the linear case $f(b_1,b_2)=b_1+b_2$ and indeed coincides with
our expression in Sorkin's arrangement. This is due to the
decomposition (\ref{proj}) for the $B$ measurement which allows us
in the linear case to transform (\ref{probm}) in equation
(\ref{prob1}). However this cannot be done in general, for
instance, in the non linear case. Furthermore there are particular
experimental setups where both formulae disagree even for the
linear case.}

\section{Conclusions}

We have developed the multi-local, covariant, relational
description of the measurement process of a quantum free field. We
have addressed the criticisms raised by various authors to the
standard Hilbert approach and shown that they are naturally
avoided by our covariant description of the measurement process.
In order to address these issues, we have extended the intrinsic
order associated to a sequence of measurements to the case of
partially connected measurement devices. This extension has
further implications on the relational meaning of the measurement
process. A particular measurement process of a given property
performed by a given measurement device on a region of space-time,
should be considered as composed by a sequence of {\it decision}
processes occurring on different regions of the device. This
solves the causal problems and implies a global relational aspect
of the complete set of alternatives $S^i$. From an observational
point of view, we have proved that causality holds in the
canonical approach for a wide and natural class of operators, while 
the standard formalism is extreemly restrictive. Our proposal could be
experimentally tested trough the implementation of the particular
configuration proposed in the previous section. Furthermore, our
predictions for the reduction of the states should be associated
to the {\it decision} process during the interaction of the
quantum system with the measurement devices and may be considered,
if confirmed, as an experimental evidence of the physical
character of the quantum states. If this is experimentally
verified, the standard {\it instrumentalist} approach introduced
by I.Bloch concerning the measurement process in relativistic
quantum mechanics, would not be compatible with experiments. This
is mainly due to the fact that this order does not coincide with
our intrinsic order in the case of partially connected regions,
and Bloch's approach would not be in general compatible with
causality for the  measurements we have considered.\\

It is now clear that the description that we have introduced has a
relational nature.\footnote{Another relational interpretation in
Q.F.T. was proposed in Ref \cite{Sud}} Firstly because the
intrinsic order of the options $S^i$ is defined in relational
terms by the measurement devices. But also because self-adjoint
operators may only be defined if they are associated to a set of
local devices. Recall that self-adjoint global operators that
describe the field on arbitrary spatial hyper-surfaces do not
exist.\\ The tendency interpretation of non-relativistic quantum
mechanics is naturally a relational theory. If one thinks, for
instance, in the solution proposed by Bohr for the EPR paradox
\cite{Bohr} one immediately recognizes that one cannot associate a
given reality to a quantum system before measurement. Even the
Unruh effect for accelerated detectors has a very deep relational
meaning. As Unruh noticed: {\em "A particle detector will react to
states which have positive frequency respect to the detectors
proper time, not with respect to any universal
time}\cite{Unruh}.\\ One of the main challenges of the XXI century
is the conclusion of the XX revolution toward a quantum theory of
gravity. The relational point of view is crucial in both theories,
the quantum and the relativistic. We have proposed a possible
interpretation for any canonical theory in the realm of special
relativity. How to extend it to gravity implies further
study, mainly because of the nonexistence of a natural intrinsic
order without any reference to a space-time background.
\footnote{This problem is connected to the meaning of
Microcausality, based only on algebraic grounds, without
background.} Furthermore, up to now there is no evidence of local
observables in pure quantum gravity. This is another evidence of
the relational character of the theory. We are now studying these
issues.

\section{Acknowledgments}

We would like to thank Michael Reisenberger for very useful
discussions and suggestions about the presentation of this paper.

\section{Appendix}

Here, we prove that the projector is a well defined operator in
the Fock space.

Let us start by looking at the quantity
$<\phi_{\Delta},t^{L_a}|0>$:

\begin{equation} <\phi_{\Delta},t^{L_a}|0>= Cdet^{1/4}(\frac{\omega}{\pi})
e^{{-i \over 2}<{\phi}_\Delta A{\phi}_\Delta>} \int D\phi e^{{-1
\over 2}<{\phi}(\omega+iA){\phi}>} e^{-i<B{\phi}_\Delta{\phi}>}
\end{equation}

Hence: \begin{equation}
<\phi_{\Delta},t^{L_a}|0>=C\frac{det^{1/4}(\frac{\omega}{\pi})}
{det^{1/2}(\frac{(\omega+iA)}{2\pi})} e^{{-i \over
2}<{\phi}_\Delta A{\phi}_\Delta>} exp\left({-1 \over
2}<{\phi}_\Delta B(\omega+iA)^{-1}B{\phi}_\Delta> \right)
\end{equation}

This turns out to be: \begin{equation}
<\phi_{\Delta},t^{L_a}|0>=\left(\prod_{p}e^{i\omega_{p}t^{L_a}
\over 2}\right) {det^{1/4}(\frac{\omega}{\pi})} e^{{-1 \over
2}<\phi_{\Delta}\omega \phi_{\Delta}>}, \end{equation}

which is just the vacuum state, evaluated for $\phi_\Delta$, up to
a global phase. This result is a consequence of the Poincare
invariance of the vacuum, modulo the zero mode, and of the fact
that in the limit when $t^{L_a}$ tends to zero
$|\phi_{\Delta},t^{L_a}>$ is just $|\phi_\Delta>$.

Furthermore, the mean value of the projector
${P^a}_{F_\Delta}(t^{L_a})$ in the vacuum state is given by:

\begin{equation} <0|{P^a}_{F_\Delta}(t^{L_a})|0>=det^{1 \over
2}\left({\omega \over \pi}\right) \int_{F_\Delta}d\Delta \int
d\phi_{\Delta}\int d\alpha e^{-<\phi_\Delta \omega \phi_\Delta>}
e^{i\alpha(<f^a \phi_\Delta>-\Delta)} \end{equation}

That is,

\begin{equation}
<0|{P^a}_{F_\Delta}(t^{L_a})|0>=\int_{F_\Delta}d\Delta\frac{2\pi}{\sqrt{<f^a
{\omega}^{-1} f^a>}} exp\left(\frac{-{\Delta}^2}{4 <f^a
{\omega}^{-1} f^a>}\right)\label{vac} \end{equation}

which leads to:

\begin{equation}
<0|{P^a}_{F_\Delta}(t^{L_a})|0>=\int_{F_\Delta}d\Delta\frac{2\pi}{\sqrt{{\left(\
\frac{2\pi}{L}\right)^3}\sum_p {\frac{|f^a(p)|^2}{\omega_p}}}}
exp\left(\frac{-{\Delta}^2}{{\left(\frac{2\pi}{L}\right)^3}{\sum}_p
{\frac{|f^a(p)|^2}{\omega_p}}}\right) \end{equation}

where $L^3$ is the volume of the box where the fields live.

Several issues may be learned from this expression. First of all,
as it should be, one gets a Gaussian distribution around the zero
value. Furthermore it is divergent free, provided the integral
$\int dp {\frac{|f^a(p)|^2}{\omega_p}}$ gives a finite result.
This is achieved by demanding that the smearing functions do not
contain high Fourier components.

In order to complete the proof we are going to show that the
projector is well defined on the complete Fock space.

To begin with, we take the single particle state:

\begin{equation} <{\phi}| 1_k>= \sqrt{2\omega_{k}}\int_{\Sigma_0}d^jy
e^{ik_jy^j}\phi(y^j) \Psi_0[\phi] \equiv
\sqrt{2\omega_{k}}\phi(k_j)\Psi_0[\phi] \end{equation}

Now we calculate $<\phi_{\Delta},t^{L_a}|1_k>$ getting:

\begin{eqnarray} &<\phi_{\Delta},t^{L_a}|1_k>=
C{det^{1/4}(\frac{\omega}{\pi})} \sqrt{2\omega_{k}} e^{{-i \over
2}<{\phi}_\Delta A{\phi}_\Delta>}{\times}& \nonumber\\ &\int
D\phi(p_j) \phi(k_j)e^{{-1 \over
2}{\left(\frac{2\pi}{L}\right)^3}\sum_p{\phi}(p_j)
({\omega}_{p}+iA(p,t^{L_a})){\phi}(p_j)}
e^{-i{\left(\frac{2\pi}{L}\right)^3}\sum_p
B(p,t^{L_a}){\phi}_\Delta(p_j){\phi}(p_j)}& \end{eqnarray}

As one immediately notices the main difference is the resulting
multiplicative factor associated with a single mode $\phi(k_j)$.

In order to include the general, many particle, case, one could
introduce a source term $J$ and define a generating functional
$Z(J)$ as follows:

\begin{eqnarray} &Z(J)=<\phi_{\Delta},t^{L_a}|0>^{J} {\equiv}
C{det^{1/4}(\frac{\omega}{\pi})}e^{{-i \over 2}<{\phi}_\Delta
A{\phi}_\Delta>}\times& \nonumber\\&\int D\phi(p_j) e^{{-1 \over
2}{\left(\frac{2\pi}{L}\right)^3}\sum_p{\phi}(p_j)({\omega}_{p}+iA(p,t^{L_a}))
{\phi}(p_j)} e^{-i{\left(\frac{2\pi}{L}\right)^3}\sum_p \left
(B(p,t^{L_a})\phi_\Delta(p_j)+J(p_j)\right){\phi}(p_j)}&
\end{eqnarray}

It is easy to show that indeed:
\begin{equation}
<\phi_{\Delta},t^{L_a}|1_k>=\sqrt{2\omega_{k}}
\frac{i\delta}{\delta J(k)} Z(J){\mid}_{J=0}
\end{equation}

This procedure may be identified with the usual one in Q.F.T., We
will call, $n$-point function the $n$ functional derivative
respect to $J(k^1)...J(k^n)$.\\ Now, it is not difficult to show
that the inner product
$<n_{1}....n_{k}...|\phi_{\Delta},t^{L_a}>$, may be calculated, up
to multiplicative factors, in terms of the $n$-point functions.
Those factors are functions of the frequencies of the modes
involved in the given Fock state.\\ To do that, we start by
studying the form of the particle states in Fock space. The Fock
space is constructed by the creation operators. Their action
applied to the vacuum in the field representation, consists in the
multiplication by some $k$ dependent component of the field
($\phi(k)$) and a derivative of the vacuum state respect to this
mode corresponding to the given particle state we were creating.
Due the structure of the vacuum, this derivative term also leads
to a multiplicative factor. It is indeed the mode $\phi_k$
multiplied by some function of the frequency of the particular
mode. Furthermore it gives a finite result since the Fock space is
made by finite set of particle states. Therefore, the Fock
$n$-particle states in the functional representation, are obtained
by multiplying the vacuum by a set of $k$-dependent components of
the field multiplied by some functions of the frequencies of each
mode in the state. This is exactly the form of the n-point
functions obtained from the generating function $Z(J)$.

The only remaining issue is the computation of $Z(J)$. One can
show that, after the functional integration is performed, one
arrives to:

\begin{equation}
Z(J)=\left(\prod_{p}e^{i\omega_{p}t^{L_a} \over 2}\right)
{det^{1/4}(\frac{\omega}{\pi})} e^{{-1 \over
2}<\phi_{\Delta}\omega \phi_{\Delta}>} e^{{-1 \over
2}<J(\omega+iA)^{-1}J>}e^{-<J(\omega+iA)^{-1}B\phi_{\Delta}>}
\end{equation}

The divergent multiplicative factor coming from the normalization
of the vacuum disappears when we take the projector as in
(\ref{vac}). Furthermore the matrix elements of the projector in
the Fock space are given by:

\begin{eqnarray}
&<n_{1}..n_{k}..|{P^a}_{F_\Delta}|n'_{1}..n'_{k}..>=&  \nonumber\\
&\int_{F_\Delta} d\Delta\int d\alpha d\phi_{\Delta}
<n_{1}..n_{k}..|\phi_{\Delta},t^{L_a}>
<\phi_{\Delta},t^{L_a}|n'_{1}..n'_{k}..>
e^{i\alpha(<f^a\phi_\Delta>-\Delta)}&
\end{eqnarray}

Since the inner product
$<n_{1}....n_{k}...|\phi_{\Delta},t^{L_a}>$ is a sum of a set of
$n$-point functions times some finite functions of the $k$-modes
of the particular state under consideration, we can write them as
derivatives of $Z(J)$ and take out of the integral the derivatives
with respect to $J_i$. Hence in the integral part it remains a
divergent factor $|{det^{1/2}(\frac{\omega}{\pi})}|\times
e^{-<\phi_{\Delta}\omega\phi_{\Delta}>}$ coming from the vacuum
state. However, the $d\phi_\Delta$ integral is quadratic in the
field and contributes with a factor that cancels this infinite, as
before. \footnote{Recall that $\int D\phi e^{<-\phi A \phi +J
\phi>}=\frac{1}{\sqrt{det(A/\pi)}}e^{\frac{1}{4}<J A^{-1}J>}$}
This is a well known fact, as it was noticed by Jackiw
\cite{Jackiw}, the divergent factor
$det^{1/4}(\frac{\omega}{\pi})=exp{V \over 4}\int d^jk {1 \over
\pi} ln\sqrt{k^2+m^2}$ , which is ultraviolet divergent, does not
affect matrix elements between states on the Fock space since it
is chosen in such a way that it it disappears from the final
expression. Thus, the matrix elements of the projector are well
defined in the Fock space and we arrive to a well defined quantum
field theory as it was required.

\end{document}